\documentclass{ansthd}
\usepackage{subcaption}
\usepackage{graphicx}
\usepackage{tabls}
\usepackage{multirow}
\usepackage[super]{nth}
\usepackage{amsmath,esint}
\usepackage[export]{adjustbox}
\usepackage{ragged2e}
\usepackage[T1]{fontenc}
\usepackage{enumitem}
\usepackage{float}
\usepackage{amsmath}
\usepackage{comment}
\usepackage{float}

\DeclareMathOperator*{\argmax}{arg\,max}
\DeclareMathOperator*{\argmin}{arg\,min}

\begin{document}

\title{Quantifying Model Uncertainty of Neural-Network based Turbulence Closures}

\author{Cody Grogan \footnote{Corresponding Author}}
\author{Som Dutta \footnote{Corresponding Author}}
\affil{Utah State University \\
  Logan, UT, 84322 \\
  A02313514@usu.edu; som.dutta@usu.edu}

\author{Mauricio Tano} 
\author{Somayajulu L.N. Dhulipala}
\affil{Idaho National Laboratory \\
  Idaho Falls, ID \\
  mauricio.tanoretamales@inl.gov; som.dhulipala@inl.gov \\\vspace{2pt}}

\author{Izabela Gutowska}
\affil{Oregon State University \\
  Corvallis, OR 97331 \\
  izabela.gutowska@oregonstate.edu}

\maketitle

\begin{abstract}
With increasing computational demand, Neural-Network (NN) based models are being developed as pre-trained surrogates for different thermohydraulics phenomena. An area where this approach has shown promise is in developing higher-fidelity turbulence closures for computational fluid dynamics (CFD) simulations. The primary bottleneck to the widespread adaptation of these NN-based closures for nuclear-engineering applications is the uncertainties associated with them. The current paper illustrates three commonly used methods that can be used to quantify model uncertainty in NN-based turbulence closures. The NN model used for the current study is trained on data from an algebraic turbulence closure model \cite{taghizadeh2021turbulence}. The uncertainty quantification (UQ) methods explored are Deep Ensembles, Monte-Carlo Dropout, and Stochastic Variational Inference (SVI). The paper ends with a discussion on the relative performance of the three methods for quantifying epistemic uncertainties of NN-based turbulence closures, and potentially how they could be further extended to quantify out-of-training uncertainties.
For accuracy in turbulence modeling, paper finds Deep Ensembles have the best prediction accuracy with an RMSE of $4.31\cdot10^{-4}$ on the testing inputs followed by Monte-Carlo Dropout and Stochastic Variational Inference. 
For uncertainty quantification, this paper finds each method produces unique Epistemic uncertainty estimates with Deep Ensembles being overconfident in regions, MC-Dropout being under-confident, and SVI producing principled uncertainty at the cost of function diversity.

\raggedleft
\textbf{KEYWORDS}\\
Uncertainty Quantification, Neural Networks, Turbulence Closure, Bayesian Neural Network, Variational Inference 
\end{abstract}

\section{Motivation} 
Thermal-hydraulic modeling in nuclear reactors is a challenging problem because it involves turbulent flow mixing at multiple scales and between reactor structures. Higher fidelity direct numerical simulations or large vortex simulations are usually prohibitively expensive at the scales involved in nuclear reactor cores. Therefore, one must rely on Reynolds-Average Navier-Stokes models or special closure correlations. However, the development of the closure equations required for these models is often complicated by the large amount of modeling required to develop them. Data-driven machine learning (ML) models are receiving significant interest for accelerating modeling and simulation tasks, including Nuclear Engineering \cite{li2022,el2021}. Within the spectrum of ML models, a class of models mathematically replicating the connection between neurons within a brain are neural networks (NN). NN-based models have the potential to provide swift and accurate prediction of complex thermohydraulic processes.
This is especially true when combining NN-based models with computational fluid dynamics (CFD) models currently in use for nuclear reactor design, analysis, and operational support.
NN based turbulence emulators
can either replace or increase the fidelity of turbulence closures used in large eddy simulations (LES) \cite{yuan2020deconvolutional} and Reynolds Average Navier Stokes (RANS) \cite{chang2020reynolds,xie2021artificial} based CFD simulations. 
The current paper illustrates the ``\textit{why and how}" of uncertainty quantification (UQ) of NN-based turbulence closures.  

One of the primary impediments to the wider adoption of these NN models from a regulatory standpoint is the lack of systematic quantification of uncertainties of these models, especially when they operate in an extrapolated state.
\cite{geneva2019quantifying} show for fluid phenomena, such as a backward step, NN-based turbulence models do not eliminate error between RANS models and their LES baselines.
\cite{geneva2019quantifying} also shows accurate uncertainty quantification of NN-base turbulence model predictions can effectively identify when a model is in an extrapolated state.
For NN-based models, an extrapolated state is when they are queried outside the range of their training data, potentially increasing the possibility of large prediction errors. This uncertainty in prediction of NN models in extrapolated state, is also know as \textit{out-of-training} uncertainty \cite{gawlikowski2023survey}. Out-of-training uncertainty of NN-models is intertwined with model/epistemic uncertainty associated with the NN models. Thus, \textbf{the objective of the current paper is to compare and contrast three different methods for quantifying model uncertainty in a NN based turbulence closure}, where the model is developed using data from an algebraic turbulence closure \cite{taghizadeh2021turbulence}. The three methods for quantifying uncertainty are Deep Ensembles, Monte Carlo Dropout, and Stochastic Variational Inference.

\section{Neural Network-based Turbulence closure}
Machine-Learning (ML) models have shown potential to significantly improve accuracy of turbulence closures used for LES and RANS based CFD simulations \cite{duraisamy2019turbulence,duraisamy2021perspectives}.
The ML models work on the rationale that the shortcomings of the physics–based closures can be mitigated by appropriate data–based training of the models. 
Among the different methods within ML, NN-based approaches have gained substantial traction, especially for RANS turbulence closure. Deep NN (DNN) with embedded invariance \cite{ling2016reynolds}, Bayesian DNN \cite{Geneva_2020} and other variations of DNN have shown promising results for different flow applications \cite{iskhakovmachine,volpiani2021machine}. 
Though, generalization of NN–based RANS modeling is difficult due to complex closure relations stemming from flow dependent non–linearity and bifurcations, and difficulty in acquiring high–fidelity data covering all the regimes of relevance. 
To get over these impediments, 
Taghizadeh et al. \cite{taghizadeh2021turbulence} proposed conducting preliminary analysis of NN closure development by  using proxy–physics turbulence surrogates.
They used algebraic Reynolds stress models (ARSM) based turbulence surrogates, which could represent the non–linearity and bifurcation characteristics observed in different regimes of turbulent flow, to generate the parameter–to–solution maps. This allowed the authors to explore the potential of using DNN to model turbulence closures across regimes, without generating large amounts of high-fidelity data \cite{taghizadeh2020turbulence}. In the current paper, our primary objective is to quantify and compare model uncertainties within NN-based turbulence closures using data from proxy-physics surrogates. 

\subsection{NN Model of Turbulence Closure using Proxy-Physics Surrogate}

In this study, we use a three–term self–consistent, nonsingular, and fully explicit algebraic Reynolds stress model (EARSM) \cite{girimaji1996fully}, with the pressure–strain correlation model proposed by Speziale, Sarkar, and Gatski (SSG) \cite{speziale1991modelling}, to generate stress-strain relationship datasets for different normalized strain and rotation rates. 
The purpose of an ARSM is to capture the characteristics of a fluid flow for different values of non-dimensional variables effectively. The details of the closure model have been illustrated succinctly. The NN models analyzed in the paper have been developed using data generated by the equations listed below. 
The closure coefficients $G_1$, $G_2$, and $G_3$ are calculated as a function of $\eta_1$ and $\eta_2$. $\eta_1 = s_{ij}s_{ij} = S^2$ and $\eta_2 = r_{ij}r_{ij} = R^2$  are the scalar invariants of the strain and rotation–rate tensors.

\begin{equation}
    G_1 = 
    \begin{cases}
        -\frac{p}{3} + \left( -\frac{b}{2} + \sqrt{D} \right)^{1/3} + \left( -\frac{b}{2} - \sqrt{D} \right)^{1/3} & D > 0 \\
        -\frac{p}{3} + 2 \sqrt{-\frac{a}{3}} \cos(\frac{\theta}{3}) & D < 0, \ b<0 \\
        -\frac{p}{3} + 2 \sqrt{-\frac{a}{3}} \cos(\frac{\theta}{3} + \frac{2 \pi}{3}) & D < 0, \ b>0
    \end{cases}
    \label{eq:G1}
\end{equation}   
\begin{equation}
    G_2 = \frac{-L_4 G_1}{L_1^0 - \eta_1 L_1^1 G_1} \quad G_3 = \frac{2 L_3 G_1}{L_1^0 - \eta_1 L_1^1 G_1}
    \label{eq:G2-3}
\end{equation}
            
Where the calculation begins with defining the SSG-related constants $C_1^0$, $C_1^1$, $C_2$, $C_3$, and $C_4$ as 3.4, 1.8, 0.36, 1.25, and 0.4 respectively. 
$L_1^0$, $L_1^1$, $L_2$, $L_3$, and $L_4$ are defined as, 
$                L_1^0 = \frac{C_1^0}{2} - 1, \quad L_1^1 = C_1^1 + 2, \quad L_2 = \frac{C_2}{2} - \frac{2}{3}, \quad L_3 = \frac{C_3}{2} - 1, \quad L_4 = \frac{C_4}{2} - 1 $
%
. The intermediate variables $p$, $q$, and $r$ are defined as, 
$                p = - \frac{2 L_1^0}{\eta_1 L_1^1}, \quad q = \frac{1}{(\eta_1 L_1^1)^2} \left[ (L_1^0)^2 + \eta_1 L_1^1 L_2 - \frac{2}{3} \eta_1 (L_3)^2 + 2\eta_2 (L_4)^2 \right], \quad r = - \frac{L_1^0 L_2}{(\eta_1 L_1^1)^2} $.
%
Finally, the intermediate values used for determining $G_1$ are,
$                a = (q - \frac{p^2}{3}), \quad b = \frac{1}{27} (2p^3 - 9pq + 27r), \quad D = \frac{b^2}{4} + \frac{a^3}{27}, \quad \theta = \cos^{-1}\left(\frac{-b/2}{\sqrt{-a^3/27}} \right) $. More details about the ARSM-SGG model used for generating the data for developing the NN can be found in Taghizadeh et al. \cite{taghizadeh2021turbulence}.

\subsection{NN Model Architecture and Performance}
The machine learning objective of this paper is to train Neural Networks to fit the ARSM data with ($\eta_1$, $\eta_2$) as inputs and ($G_1$, $G_2$, $G_3$) as outputs.
To generate training, validation, and testing data, each input point is generated at random uniformly in the log base 10 space.
Generating training and test points in the log space ensures areas of high change are represented adequately in the training data as a majority of the change in ($G_1$, $G_2$, $G_3$) happens when the inputs are between zero and 10.
This paper uses 80,000 training data points for training, 40,000 points for validation, and a 700 by 700 grid (490,000 points) for testing.
As the true function of the inputs is known, a large testing dataset is generated to ensure the accuracy of each Network is tested in a robust manner that approximates the true prediction error.
After generating the data, each of the inputs and outputs are transformed to simplify the training and increase accuracy.
For the inputs, $\eta_1$ and $\eta_2$ are transformed by applying a natural logarithm and then using Sci-Kit Learn's standard scaler ($\eta^* = \frac{\ln(\eta) - mean(\ln(\eta))}{std(\ln(\eta))}$)
For the outputs, the negative G values are negated, a natural logarithm transform is performed, and then each output is scaled using Sci-Kit Learn's standard scaler ($G^* = \frac{\ln(|G|) - mean(\ln(|G|))}{std(\ln(|G|))}$).
Scaling in this way has the effect of shrinking the areas of low change in the ARSM and reducing the complexity of the function.
In practice, this transformation reduces the necessary model parameters from nearly 40,000 to just under 1,000. 
The loss of each method is reported in root mean squared error (RMSE) to promote the interpretability of the error.

The Neural Network in this paper is a Feed-Forward network, as it is the most general, with 4 total layers (1 input, 2 hidden, 1 output) with a hidden node size of 20 (963 learnable parameters) and uses the adaptive moment estimation (ADAM) version of SGD for optimization.
Each layer uses a ReLU activation function except for the last where a linear transformation is used because the outputs need to have varying scales and the ability to be negative.
Before addressing the specific hyper-parameters for each method, it is important to note the motivation of this paper is uncertainty quantification and not model accuracy.
As a result, a rough grid search is done over the parameters of each method to ensure the performance of each method is approximately optimal.
The \textbf{Deep Ensemble} made up of 40 members trained with a weight regularization of $10^{-7}$, batch size of 64, a decaying learning rate (successive halving) starting at 0.001, and Mean Squared Error (MSE) loss.
The \textbf{Monte-Carlo Dropout} network is trained with dropout percentage of 0.1\% on all layers except the output, a batch size of 64, a decaying learning rate (successive halving) starting at 0.001, and Mean Squared Error (MSE) loss.
The \textbf{Stochastic Variational Inference} network is trained with a batch size of 512, a Low-Rank Multivariate Gaussian prior over the parameters, a constant learning rate of 0.001, and a Multivariate Gaussian likelihood with diagonal covariance ($\sigma^2 = 0.1$).

\begin{figure}[h]
    \centering
    \includegraphics[scale=0.3]{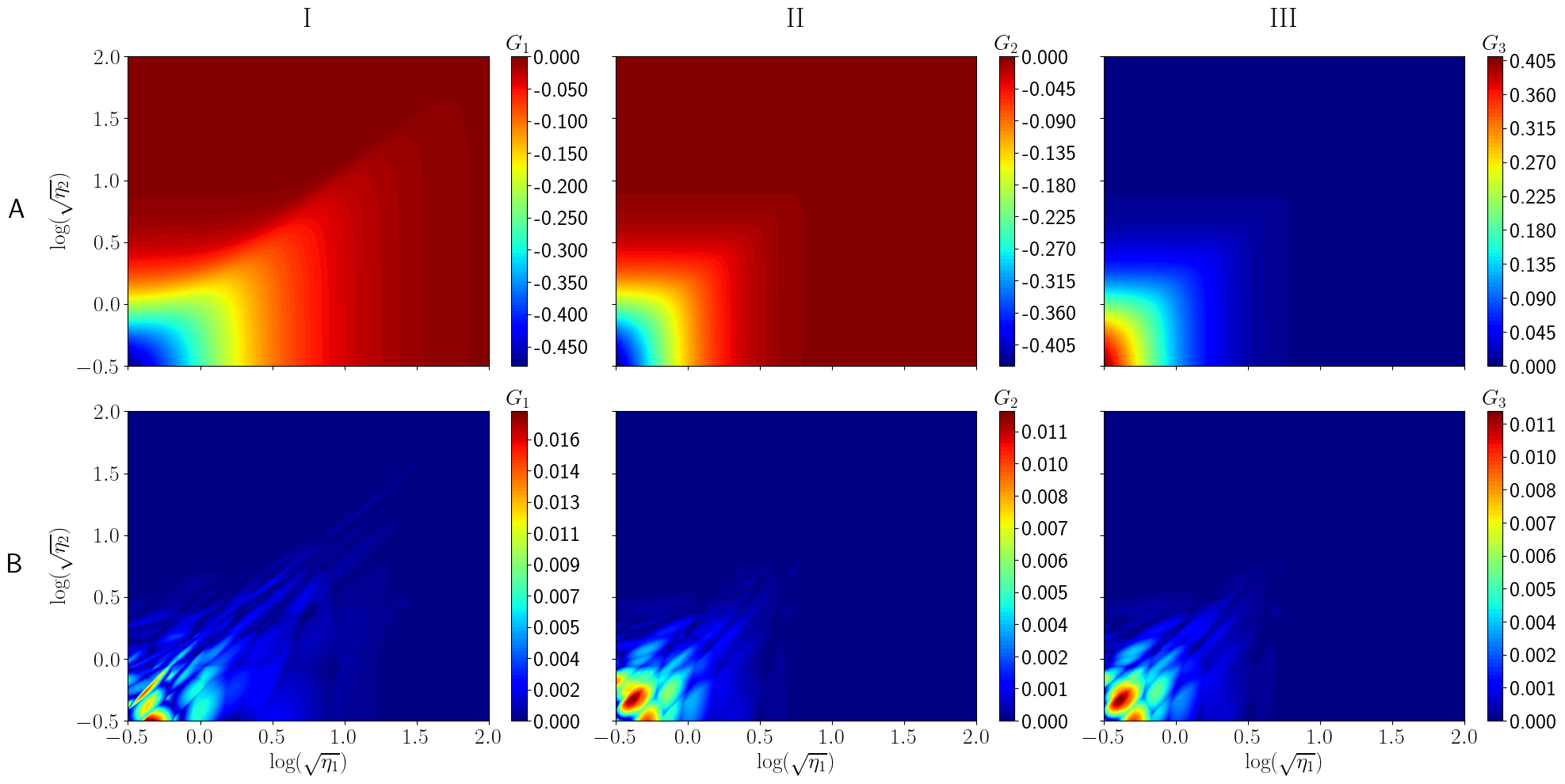}
    \caption{A plot of the true ($G_1$, $G_2$, $G_3$) from the ARSM (row A) and absolute error of Neural Network predictions (row B) with a test loss of $6.80\cdot 10^{-4}$ RMSE and $7.18\cdot 10^{-3}$ MAPE.}
    \label{fig:Deterministic_Preds}
\end{figure}

To show the sufficient complexity of the NN structure, Fig. \ref{fig:Deterministic_Preds} row B shows the accuracy of a single deterministic network with identical training to the Deep Ensemble networks.
As one can see, the prediction plots in row B are nearly indistinguishable from the reference plots in row A, which is reinforced by the RMSE of $6.80\cdot 10^{-4}$.
This shows the proposed network structure is sufficient to predict the turbulence closure parameters from the ARSM accurately.
In addition, comparing the results, the RMSE and MAPE testing errors of the proposed Neural Network are an order of magnitude lower than the results in \cite{taghizadeh2021turbulence}.
This likely stems from the transformations applied to the inputs and outputs of the problem, which enable small networks to achieve better accuracy by shrinking areas of low significance.

\section{UQ of the NN Turbulence Closure Model}
\subsection{Fundamentals of NN Uncertainty Quantification}
A Neural Network is an operation on some input defined by repeatedly applying some vector-valued function followed by a non-linear transformation.
The simplest Neural Network is called a Multi-layer Perceptron or Feed-Forward Network. 
In this Neural Network, the vector-valued function is a linear transformation with a bias followed by an element-wise non-linearity $\textbf{o}_i = \sigma(W_i \textbf{o}_{i-1} + b_i)$.
Where $\textbf{o}_i$ is the output from layer i, $W_i$ is the weight matrix for layer i, $b_i$ is a bias vector for layer i, and $\sigma$ represents the element-wise non-linearity.
Repeated evaluation of the base function yields a Feed-Forward network with L layers (if it is evaluated L times) with $\textbf{o}_0$ being a vector of inputs or observed features in data.
The number of layers in a Neural Network is often referred to as a hyper-parameter of the learning.
Another important hyper-parameter is the non-linearity $\sigma$, which is typically the Sigmoid, ReLU, Leaky ReLU, or Hyperbolic tangent function.
Other hyperparameters include optimization algorithm, regularization, dimension of layers (projection of input to higher or lower space), and the loss function.
    
As one can already see, there are a significant number of user-defined parameters that must be chosen before attempting to learn a Neural Network function from a dataset.
The learning of a Neural Network is simply selecting the weights and biases for each layer that minimizes some metric of inaccuracy or loss on a given dataset.
\begin{equation}
    \theta^* = \argmin_{\theta} L(y, \hat{y})
    \label{eqn:NN_obj}
\end{equation}
Where $\theta$ represents all of the weights and biases of the Neural Network, $L$ is the loss function, $y$ is a vector of the true outputs, and $\hat{y}$ is a vector of the predicted output of the NN for each input.
Eq. \ref{eqn:NN_obj} shows that a NN aims to find parameters $\theta$ that minimize loss, which is typically non-convex, typically using Stochastic Gradient Descent (SGD).
A non-convex loss means more than one minimum in the loss function exists or more than one set of Neural Network parameters satisfies it.
Some sets of Neural Network parameters may be algebraically equivalent, multiple Neural Network parameters create the same underlying function, or the parameters may underly entirely different functions.
As a result, one can only conclude that a Neural Network learns an estimate of the function generating the data, where hyperparameter selection affects the complexity and form of proposed functions.

The uncertainty of a Neural Network can be broken into Aleatoric (Data) and Epistemic (Model) uncertainty.
Aleatoric uncertainty or data uncertainty is the uncertainty in the dataset or noise in the observations on which one wants to train the network.
Epistemic or model uncertainty is the uncertainty in the learned function to predict the data or in this paper the Neural Network.
However, a more intuitive description of these uncertainties is irreducible (Aleatoric) and reducible (Epistemic) uncertainty.
Aleatoric uncertainty is irreducible because additional data points from the same noisy observer can not reduce the noise in the data.
Epistemic uncertainty is reducible because additional data points outside the current input space can reduce the number of potential functions that predict the data.



In statistics, the idea that there is one set of optimal parameters for Eq. \ref{eqn:NN_obj} is called frequentist.
However, the previous section highlights how there are likely multiple Neural Network parameters that achieve satisfactory accuracy on a given dataset.
This idea that multiple solutions exist is referred to as Bayesian statistics.
\cite{GoodBengCour16} defines the difference between these approaches as frequentists believe the dataset is random and Bayesians believe the parameters of a model are random.
The Bayesian approach allows for the Epistemic uncertainty to be modeled through what is called the predictive distribution.
\begin{equation}
    p(y|x,D) = \int_{\Theta} p(y|x,\theta)p(\theta | D) d\theta
    \label{eq:pred_dist}
\end{equation}
Eq. \ref{eq:pred_dist} can be read as the probability distribution of an output given an input and a dataset.
The right-hand side of Eq. \ref{eq:pred_dist} is a weighted average over all possible parameters $\theta$ of the probability of an output given an input and parameters.
More intuitively, the predictive distribution characterizes the possible values y can take on for all possible models that map x to y where models that better fit the dataset get more influence.
For those more familiar with probability, one might recognize Eq. \ref{eq:pred_dist} as an expectation of $p(y|x,D)$ over $p(\theta|D)$.
This allows for the use of Monte-Carlo integration, also referred to as model averaging, to get an approximation of the predictive distribution.
\begin{equation}
    p(y|x,D) \approx \frac{1}{N} \sum_{\theta_i \in \Theta} p(y|x, \theta_i) \quad \theta_i \sim p(\theta |D) 
    \label{eq:MC-pred}
\end{equation}
Using Eq. \ref{eq:MC-pred} one can calculate the characteristics of $p(y|x,D)$ with a finite number of parameter samples from $p(\theta |D)$.
For instance the mean and covariance in regression are,
\begin{equation}
    \hat{y} = \frac{1}{N} \sum_{\theta_i \in \Theta} f_{\theta_i}(x) \label{eq:mean}
\end{equation}
\begin{equation}
    \Sigma = \frac{1}{N-1} \sum_{\theta_i \in \Theta} (f_{\theta_i}(x) - \hat{y})^T(f_{\theta_i}(x) - \hat{y}) \label{eq:cov}
\end{equation}
Although this may seem like a simple process, in practice, it is difficult to generate accurate samples from $p(\theta |D)$.
$p(\theta |D)$ is what is known as the posterior distribution of $\theta$ and its order of conditioning on the data is not easy to evaluate.
However, one can express the posterior distribution in a more evaluatable way using Baye's Rule.
\begin{equation}
    p(\theta |D) = \frac{p(D|\theta) p(\theta)}{p(D)} = \frac{p(D|\theta)p(\theta)}{\int p(D | \theta)p(\theta) d\theta}
    \label{eq:posterior}
\end{equation}
Thanks to Baye's rule, one can interpret the posterior distribution of $\theta$ as the multiplication of the probability of the parameters explaining the data and the probability of the parameters occurring.

The probability of the parameters explaining the data is often referred to as the likelihood of the data given some parameters $\theta$.
For Neural Networks, this likelihood is related to the loss of a Neural Network on the training data or Neural Networks with higher accuracy have a higher probability of explaining the data (one can also show the minimizing MSE loss is the same as maximizing the Gaussian likelihood of the data).
The probability of the parameters $\theta$ in Eq. \ref{eq:posterior} is referred to as the 'prior' distribution or how one incorporates prior knowledge into the distribution.

The final term, and most difficult to evaluate, is the denominator in Eq. \ref{eq:posterior} and is the probability of the data over all possible parameters.
As one can already see, the integral for any Neural Network is impossible to evaluate analytically over the entire parameter space.
However, this integral evaluates to a constant or acts as a normalizer, which means the posterior distribution has the same 'shape' as the numerator in Eq. \ref{eq:posterior}.
Because it is a constant, researchers have devised ways to get around evaluating the integral and generate parameter samples from the posterior so they can be used to approximate the predictive distribution.

\subsection{Deep Ensembles}

In machine learning, an ensemble is created by combining the knowledge of multiple learners by taking a weighted sum of their outputs.
The proportion of the contribution can be different in the case of boosting, or the same, which facilitates the parallel training of the learners.
For more traditional ensembling methods, each member of the ensemble must be randomized using data or other factors to promote functional diversity among the members.
For Neural Networks, \cite{Deep-Ensembles,fort2020deep} show the random initialization of the parameters and shuffling of data is sufficient to promote functional diversity among members of an ensemble.
The use of ensembles is especially useful when a model class has low bias and high variance, which allows averaging to reduce the variance of predictions.
A more mathematically rigorous notion is the posterior distribution of a Deep Ensemble is approximated as the mixture of delta functions.
\begin{equation}
    p(\theta |D) = \frac{1}{N}\sum_{i=1}^N \delta(\theta - \theta_l)   
    \label{eq:ensemble_post}
\end{equation}
Eq. \ref{eq:ensemble_post} shows each ensemble member has an equal likelihood of occurring and the posterior has point densities at the parameters of each ensemble member.

Although a single deterministic network is frequentist, \cite{gordon2021} shows an Ensemble of deterministic networks is equivalent to Bayesian model averaging through Eq. \ref{eq:MC-pred}.
This requires showing SGD on a randomly initialized Neural Network generates a valid sample from the posterior.
One can show the optimization objective in Eq. \ref{eqn:NN_obj} for SGD is equivalent to generating a MAP (Maximum A-Posteriori) estimate from the posterior with a uniform prior over the parameters.
Where a MAP estimate of the parameters is a set of parameters that maximizes the posterior over the weights.
This allows SGD to sample from the different likelihood peaks in the posterior making Deep Ensembles potentially 'Multi-Modal' or can characterize multiple high-probability parameters that may be entirely different functions.
The main disadvantage of Deep Ensembles is its linear scaling of training time, compute resources for inference, and long-term storage.
There exist ways to circumvent these disadvantages such as large cyclic learning rates to generate multiple ensemble members from a single training, at the cost of reducing function diversity.

\begin{figure}[h]
    \centering
    \includegraphics[scale=0.3]{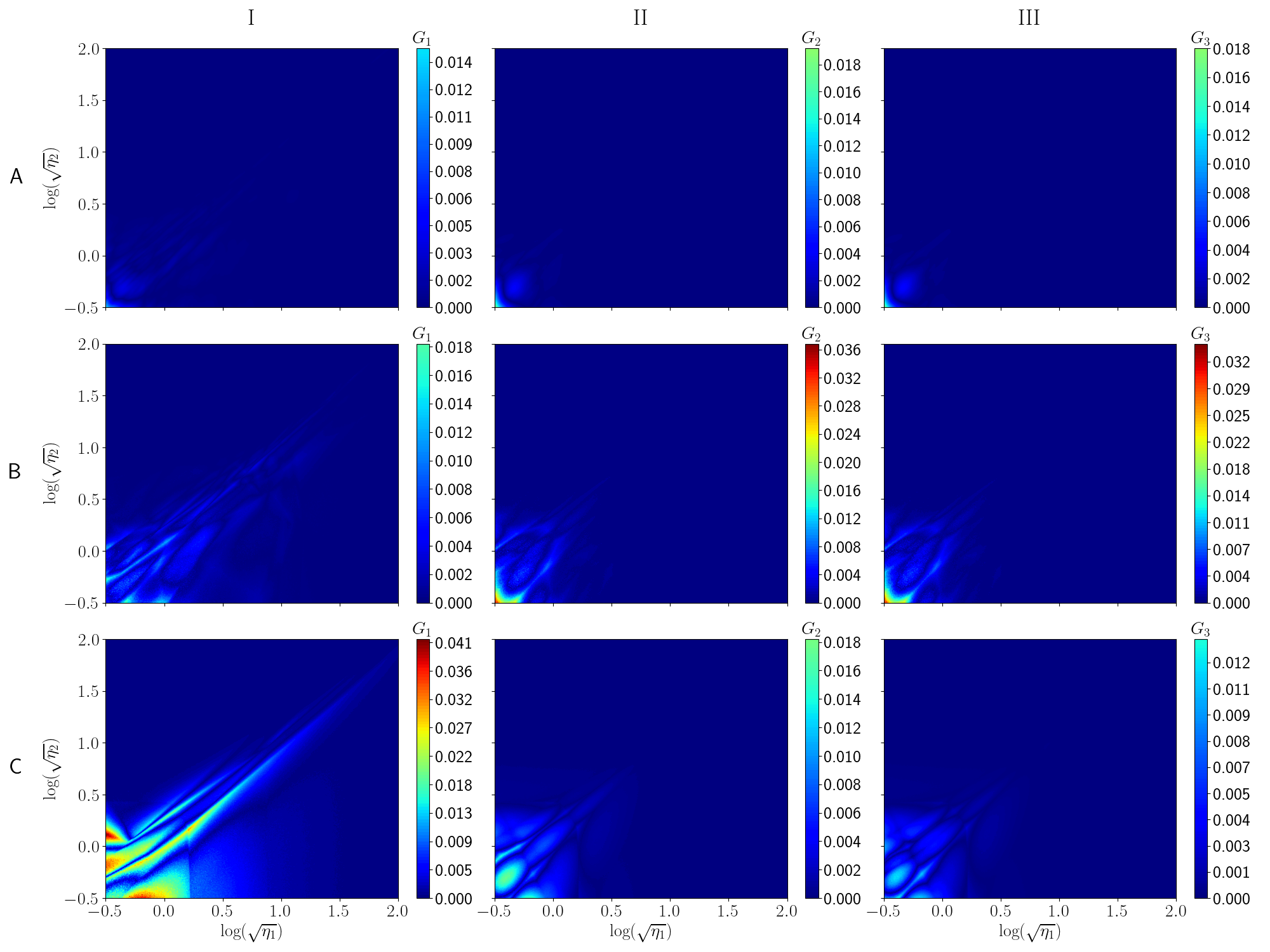}
    \caption{A plot of the absolute error between predictions and true values of ($G_1$, $G_2$, $G_3$) as a function of inputs for an Ensemble of 40 networks (row A, RMSE $4.31\cdot 10^{-4}$), Monte-Carlo Dropout (row B, RMSE $1.24\cdot 10^{-3}$), and Stochastic Variational Inference (row C RMSE, $3.35\cdot 10^{-3}$).}
    \label{fig:Pred_diff}
\end{figure}

Fig. \ref{fig:Pred_diff} row A shows the difference between the prediction of the Deep Ensemble and the true values of ($G_1$, $G_2$, $G_3$).
The ensemble exhibits high accuracy because the prediction of the ensemble averages predictions over multiple highly probable Neural Networks.
This shows ensembling reduces the variance of each of the Neural Networks or collectively allows them to eliminate small errors in predictions.
Fig. \ref{fig:STD} row A shows the standard deviation of the predictions and quantifies the Epistemic uncertainty of the proposed Neural Network structure and training data.
This effectively shows how confident the ensemble is in its prediction and quantifies how much the ensemble member's predictions differ from one another.
For example, the lower left of the $G_2$ plot shows an error of roughly 0.018 where the corresponding spot in the standard deviation plot has a value of 0.006.
By dividing the error by the deviation, it can be seen that the true value of a prediction in this area is approximately 3 standard deviations away or the network is 99\% sure the true value is erroneous.
This shows Deep Ensembles only have the potential to take on entirely different functions and do not guarantee accurate uncertainty estimates.

\subsection{Monte-Carlo Dropout (MC-Dropout)}
Monte-Carlo Dropout is widely used as a regularization technique for large networks to improve generalization.
Monte Carlo Dropout consists of adding a diagonal matrix after the application of the non-linearity, where the entries are generated by a Bernoulli distribution $z_{ii} \sim Bernoulli(p)$.
These Dropout matrices are then generated for each prediction of the NN which 'drops' p \% of the elements in an output vector (nodes) for a given layer of a Neural Network.
In the limit of an infinitely wide NN, \cite{gal2016dropout} shows any network structure with Dropout approximates a Deep Gaussian Process.
Where a Gaussian process has an infinite number of parameters where any finite number of parameters has a multivariate normal distribution.
In the same way, although not infinite, a Neural Network trained with Monte-Carlo Dropout is a collection of parameters where any arbitrary subset is a function that approximates the data.
This means the training of the network creates the posterior in the form of a Neural Network and one can generate samples from the posterior by randomly dropping out p\% nodes.
    
The main question about MC-Dropout is whether it is a multi-modal or a single-modal approximation of the posterior.
As each random draw is a different function that approximates the data, one may assume MC-Dropout is multi-modal.
However, for a finite number of parameters, each function drawn must be a subset of the Neural Network parameters, which produces functions with similar parameters.
This produces inherent coupling between the different function realizations using MC Dropout and likely results in a single modal posterior.
The keyword here is 'likely single-modal' because a large Neural Network with a large dropout rate can potentially produce functions that are not coupled.
The benefit of MC-Dropout is that it circumvents the necessity of storing multiple unique networks, which reduces training time and storage.
However, the benefits come at the cost of introducing new hyper-parameters, a tendency toward single-modal approximations, and reducing model expressiveness.

\begin{figure}[h]
    \centering
    \includegraphics[scale=0.3]{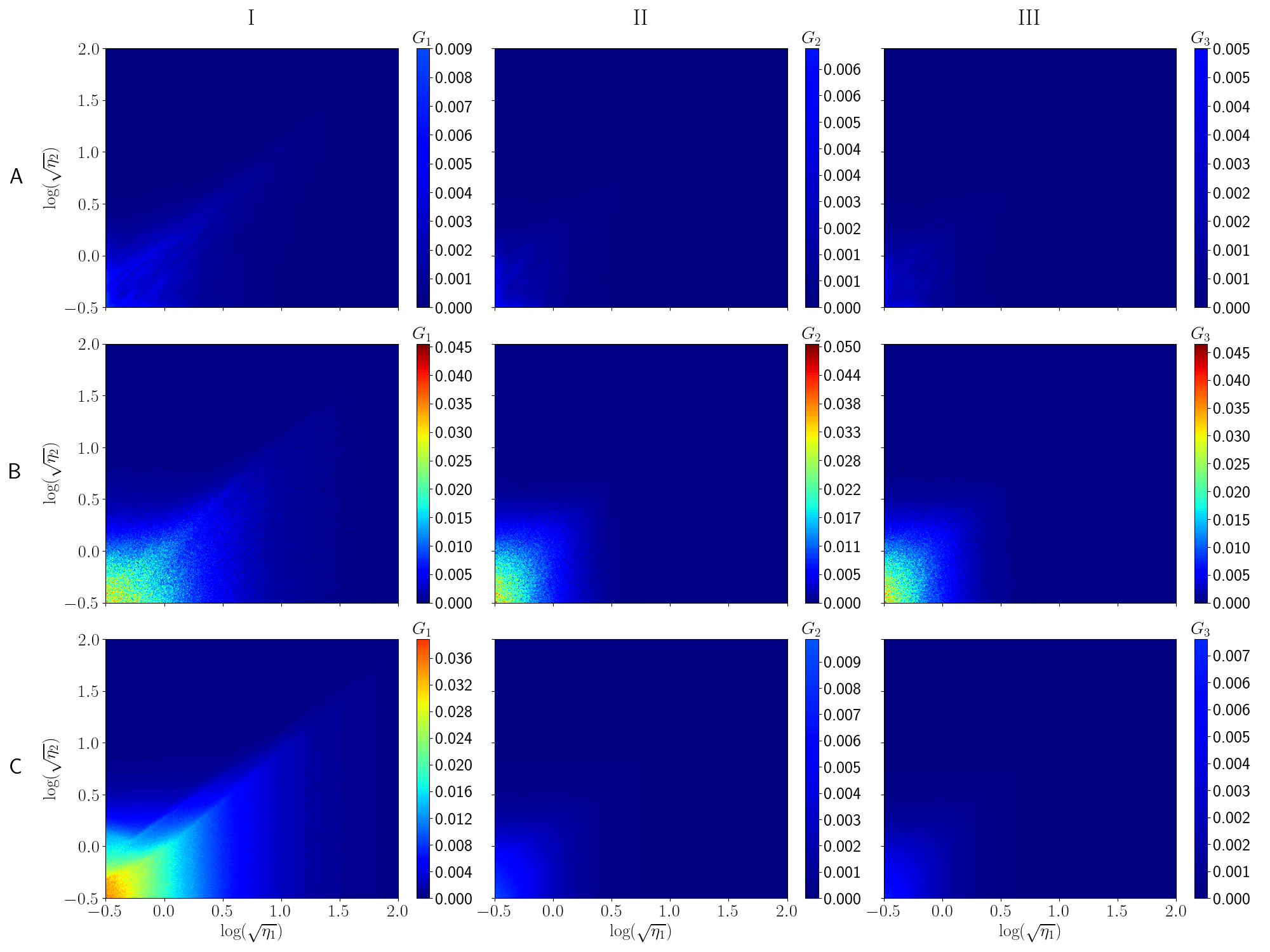}
    \caption{A plot of the standard deviation (square root of the diagonal of the covariance matrix) of the prediction from an Ensemble of 40 networks (row A), Monte-Carlo Dropout (row B), and Stochastic Variational Inference (row C).}
    \label{fig:STD}
\end{figure}

Fig. \ref{fig:Pred_diff} row B shows the accuracy of Monte-Carlo Dropout is slightly worse than the Deep Ensemble of 40 networks.
This stems from MC-Dropout randomly reducing the complexity of the Neural Network by p\%, which reduces the complexity of the proposed function.
In addition, it is not guaranteed a randomly chosen set of parameters is optimized to minimize the loss in all regions.
Meaning, it is likely a specific configuration of the network parameters has not learned enough about how to predict a specific area of the input.
Fig. \ref{fig:STD} row B shows the epistemic uncertainty of 150 predictions using different sets of network parameters.
As one can see, the standard deviation is significantly higher than the other methods, and the error of its predictions.
This reinforces the conclusion the MC-Dropout training process creates more variance in the learned functions.
However, this result likely does not extend to all problems because the number of parameters for the network is relatively low.
For instance, Neural Networks with many more parameters than the number of data points would likely have lower variance.

\subsection{Stochastic Variational Inference (SVI)}

Stochastic Variational inference is yet another way of producing a posterior distribution over the weights of a Neural Network given some data.
In general, SVI is an extension of standard variational inference where the gradients are estimated from subsets of the data \cite{hoffman2013stochastic}.
The key part of this sampling technique is defining a variational distribution which greatly simplifies some of the computation for inference.
The optimization algorithm then maximizes the Evidence Lower Bound (ELBO) using SGD, which minimizes the KL divergence between the variational posterior and true posterior. \\
\begin{equation}
\argmin_{q_\phi} D_{KL}(q_\phi(\theta) || p(\theta | D)) \approx \argmax_{q_\phi} ELBO(q_\phi(\theta) || p(D|\theta)p(\theta)) 
\label{eq:ELBO}
\end{equation}
In simpler terms, the objective of Eq. \ref{eq:ELBO} is to make some simpler distribution $q_\phi$ and its parameters $\phi$ (simple to sample) very similar to the posterior $p(\theta|D)$ (intractable to sample).
Through the ELBO in Eq. \ref{eq:ELBO}, one can relate similarity to the posterior as similarity to the numerator of the posterior, avoiding the intractable denominator.
This paper employs a Multivariate Gaussian variational distribution with a low-rank covariance matrix to estimate the posterior over the Network parameters.
This results in a Network that approximates a single mode or function of the posterior with a Gaussian distribution in very high dimensional space.
Samples from this simpler distribution can then be generated to quantify the uncertainty of the predictive distribution.
Where each sample from the approximate posterior distribution slightly perturbs the Neural Network's learned function.

The main benefit of using SVI is its ability to scale Bayesian Inference to large networks in a tractable amount of training time.
The disadvantages of SVI are a lack of functional diversity, a difficult optimization objective, and the limitations of the variational distribution.
As mentioned previously, SVI only learns the uncertainty in the parameters of a Neural Network around a single function, which can lead to low-quality uncertainty estimates.
Another limitation to SVI is the ELBO optimization objective is a more abstract and difficult quantity to calculate, which results in additional noise in the learning and thus degrades accuracy compared to MSE loss.
The final limitation of SVI is the proposed variational distribution over the parameters of a Neural Network limits their interaction.
This means assuming some structure on the interaction between some parameters in the Neural Network may end up reducing the complexity of the functions a network trained with SVI can represent.

Fig. \ref{fig:Pred_diff} row C shows the prediction error for a Neural Network trained with SVI and a Low-Rank Gaussian prior.
As previously mentioned, the Low-Rank Gaussian Prior and the difficult objective function greatly reduce the accuracy of SVI compared to the other methods.
In addition, the SVI method is only learning the uncertainty of the parameters of a single function meaning it is susceptible to the variance of the learning.
Whereas Ensembles and MC-Dropout reduce the variance of their prediction by averaging over multiple functions, increasing their accuracy.
Fig. \ref{fig:STD} row C shows the standard deviation of SVI is significantly lower than the MC-Dropout Network.
Most interesting is the SVI network has significantly different standard deviations for $G_1$, $G_2$, and $G_3$.
Suggesting the peaks in probability around the parameters related to $G_2$ and $G_3$ are likely narrower than peaks around parameters for $G_1$.
In other words, the function produced by the network trained with SVI is likely expressive enough to calculate $G_2$ and $G_3$, but not expressive enough to calculate $G_1$.
This suggests that increasing the Neural Network's size could potentially increase the accuracy of the network, as is expected.
In addition, it is important to emphasize that SVI is only learning the uncertainty of the parameters of a single function rather than many.
The effect of only learning a single function is apparent in the standard deviation of the SVI network's lower left prediction of $G_2$ and $G_3$. 
Where the true value of the prediction is nearly 2 standard deviations away from the SVI network's prediction and indicates the network believes the true value is improbable.

\section{Discussion}
To begin the discussion of the different uncertainty quantification methods for Neural Networks, this paper gives an additional, more qualitative plot of the different methods.
To do this, one must think of a Neural Network in its most fundamental form, an operator that maps real-valued inputs to real-valued outputs.
Although the mapping of a Neural Network is defined on an infinite input space, one can characterize the function or posterior sample of a Neural Network using a finite number of inputs.
Each posterior sample can then be visualized using dimensionality reduction methods such as PHATE (Potential of Heat-diffusion for Affinity-based Trajectory Embedding) \cite{moon2019phate}.
A method such as PHATE balances the influence of global and local structure in the reduced data to produce visualizations with more meaningful distance and directional relationships.

\begin{figure}[H]
    \centering
    \includegraphics[scale=0.5]{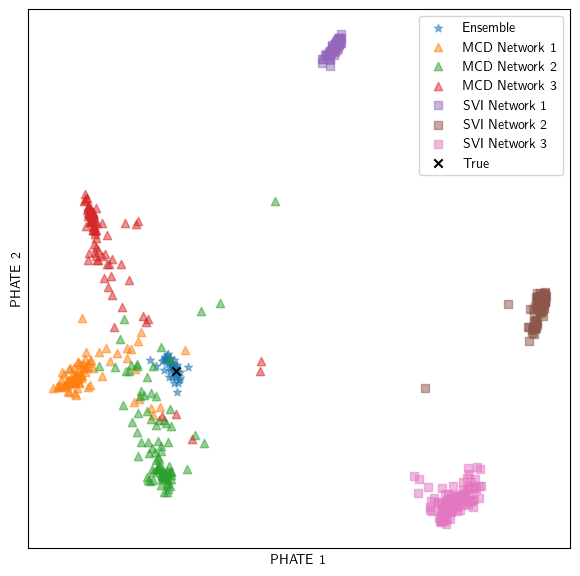}
    \caption{A plot of the PHATE dimensionality reduction embedding of 900 test predictions for each Ensemble member, 3 MC-Dropout Networks, and 3 SVI networks}
    \label{fig:PHATE}
\end{figure}

Fig. \ref{fig:PHATE} shows the posterior approximation samples of an Ensemble, three different MC-Dropout Networks, and three different SVI Networks.
Analyzing the global structure of the plot, the accuracy hierarchy of different methods is shown by the relative distance from any point to the reference point.
This shows qualitatively Ensembles have the highest accuracy followed by MC-Dropout and SVI.
Analyzing the local structure in Fig. \ref{fig:PHATE}, one can relate the Epistemic uncertainty or spread of the Turbulence closure surrogates to prediction confidence.
One can see the small spread or relatively low Epistemic uncertainty in the predictions of the Deep Ensemble samples.
This indicates a surrogate using Deep Ensembles for uncertainty quantification will be very confident, sometimes overconfident, in its predictions of flow parameters ($G_1,G_2,G_3$).
Conversely, one can see the large spread or large uncertainty of the MC-Dropout samples.
This means a surrogate using MC-Dropout for uncertainty quantification will often be underconfident in its predictions, which may lead to a modeler unnecessarily discarding an accurate prediction because of high uncertainty.
In another interesting result, one can see the different SVI networks have high error and relatively low uncertainty.
This may allow a modeler to use an inaccurate prediction because the Epistemic uncertainty suggests the prediction is confident.
Fig. \ref{fig:PHATE} illustrates pitfalls of Epistemic uncertainty quantification in NN-based surrogates using approximate methods such as the ones in this paper.
This laments the importance of accurate Epistemic uncertainty quantification because, with accurate estimates, a modeler can revert to a principled model, such as $k$-$\omega$, in the presence of high surrogate uncertainty.
\section{Conclusions}
%
This paper has presented different methodologies for Epistemic uncertainty analysis of turbulence closures in CFD models.
We use data generated from an Algebraic Reynolds Stress Model (ARSM) for turbulence closure to train NN models, and illustrate the benefits and limitations of Deep Ensembles, Monte-Carlo Dropout, and Stochastic Variational Inference in quantifying epistemic uncertainty of the NN models. The main conclusions of this study are the following:
\begin{enumerate}
    \item The Deep Ensemble produces the highest accuracy of the three methods but produces overconfident uncertainty estimates due to a lack of functional diversity.
    \item Monte-Carlo Dropout produces the second-highest accuracy but, because of the noise from introduction of Dropout, it is under-confident in its predictions.
    \item Stochastic Variational Inference, performs worst in accuracy. It gives principled uncertainty estimates around its learned function, but its estimates are not guaranteed to be accurate as it does not model multiple functions.
\end{enumerate}
Future work will focus on using similar methods to quantify out-of-training uncertainty, which is the uncertainty of inputs outside of the training dataset. This will allow for the characterization of predictions in which the model has little to no knowledge of what the prediction should be. Additionally, this study will be extended to include quantification of aleatoric uncertainty. In the future, the UQ analysis will be extended to deeper and larger NN models of turbulence closures, trained on turbulence data from high-fidelity LES and RANS based CFD simulations of Nuclear Engineering related thermohydraulics phenomena.


\section*{ACKNOWLEDGMENTS}
This research has been funded by DOE through the NEUP program, under the grant DE-NE0009397.

\setlength{\baselineskip}{12pt}

\bibliographystyle{ansthd}
\begingroup
\raggedright
\bibliography{references}
\endgroup




\end{document}